# A Pure Integral-Type PLL with a Damping Branch to Enhance the Stability of Grid-Tied Inverter under Weak Grids

Yi Zhou, Zhouchen Deng, Shi Chen, Yiwei Qiu, Tianlei Zang, Buxiang Zhou

*Abstract*—In a phase-locked loop (PLL) synchronized inverter, due to the strong nonlinear coupling between the PLL's parameters and the operation power angle, the equivalent damping coefficient will quickly deteriorate while the power angle is close to 90° under an ultra-weak grid, which causes the synchronous instability. To address this issue, in this letter, a pure integral-type phase-locked loop (IPLL) with a damping branch is proposed to replace the traditional PI-type PLL. The equivalent damping coefficient of an IPLL-synchronized inverter is decoupled with the steady-state power angle. As a result, the IPLL-synchronized inverter can stably operate under an ultra-weak grid when the equilibrium point exists. Finally, time-domain simulation results verify the effectiveness and correctness of the proposed IPLL.

## I. INTRODUCTION

THE small-signal synchronous instability issues in PLL-synchronized inverters under the weak grid have been discussed in many literature, which indicate that the main reason for instability is the negative damping dynamic caused by the PLL. Under the low short-circuit ratio (SCR) of a very weak grid, the damping ratio sharply drops [1].

At present, to enhance the synchronous stability of PLL-synchronized inverters under the weak grid, some solutions including parameter optimization and structural optimization are proposed. In [2], the parameters tunning methods of PLL and current loop are proposed to keep the stability under SCR=1. However, the effectiveness of parameter tuning is limited under uncertain grid conditions. To improve the robustness of PLL-synchronized inverters under uncertain grid conditions, some improved control structures are also proposed, such as additional inertia and damping loops in PLL[3], additional filters in PLL[4], double-PLLs-based inverters [5] and inner current loop

The authors:Yi Zhou, Zhouchen Deng, Shi Chen, Yiwei Qiu, Tianlei Zang, and Buxiang Zhou are with the School of Electrical Engineering, Sichuan University, Chengdu 610065, China. (E-mail: zhouyipower@163.com, dengzhouchen@stu.scu.edu.cn, chen_shi@scu.edu.cn, ywqiu@scu.edu.cn zangtianlei@scu.edu.cn, zhoubx@scu.edu.cn).

The corresponding author is Yiwei Qiu. Currently, he is an Associate Research Fellow at Sichuan University, Chengdu 610065, China. E-mail: ywqiu@scu.edu.cn.

reshaping method[6]. Additionally, based on the grid impedance online measurement, [7] proposed an impedance compensation method to decouple the impacts of grid impedance on stability. However, these modified PLLs or current control structures are complex.

In this letter, it is found that the equivalent damping coefficient of a PLL-synchronized inverter will quickly deteriorate when the power angle is close to 90° under an ultra-weak grid due to the strong nonlinear coupling between PLL's parameters and operation power angle, even if the PLL's parameter is well tuned under normal operation conditions. Therefore, in the proposed IPLL, the proportional loop in the traditional PI-type PLL is removed, and a damping feedback loop is added to the IPLL to keep it stable. As a result, the equivalent damping coefficient of an IPLL-synchronized inverter is decoupled with the steady-state power angle. The IPLL-synchronized inverter can keep good damping performance even in an ultra-weak grid. In addition, like a virtual synchronous machine, its equivalent inertia and damping coefficients can be also independently designed, respectively.

## II. MODELING OF TRADITIONAL PLL-SYNCHRONIZED INVERTER

The circuit and control structure of the traditional PLL-synchronized inverter is shown in Fig. 1.

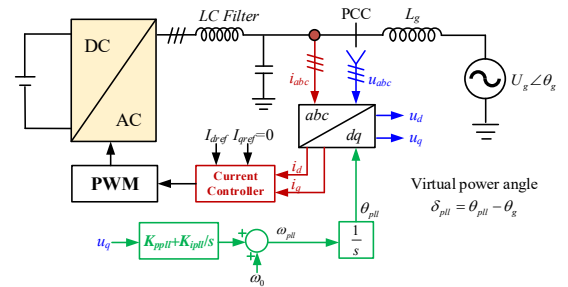

Fig. 1. Circuit and control structure of a traditional PLL-synchronized inverter.

Since the dynamic of the current loop is faster than PLL, the impacts of the current loop in analyzing the synchronous stability are not considered. Moreover, in a PLL-synchronized inverter, the virtual power angle $\delta_{pll}$ is defined as the difference between the phase angle of PCC's voltage and the phase angle of grid voltage. As a result, the second-order synchronization equation of a traditional PLL-synchronized inverter is written as follows.

$$\begin{cases} \dot{\delta}_{pll} = \omega_{pll} - \omega_g \\ \underbrace{\frac{1 - K_{ppll} L_g I_{dref}}{K_{ipll}}}_{J_e} \dot{\omega}_{pll} = \underbrace{\omega_g L_g I_{dref} + R_g I_{qref}}_{P^*} - \underbrace{U_g \sin \delta_{pll}}_{P} \\ \quad - \underbrace{\left[ \underbrace{\frac{K_{ppll} U_g}{K_{ipll}} \cos \delta_{pll}}_{D_{e1}} \underbrace{- L_g I_{dref}}_{D_{e2}} \right]}_{D_e} (\omega_{pll} - \omega_g) \end{cases} \quad (1)$$

In (1), $J_e$ represents the equivalent inertia coefficient, $D_e$ represents the equivalent damping coefficient. From (1), it can be seen that both equivalent inertia and damping coefficients vary under different grid impedance and operation currents. With increasing grid impedance and operation current, the equivalent inertia coefficient is decreased. Also, the equivalent damping coefficient of a traditional PLL-synchronized inverter is indefinite [8]. Under the inverter mode, $D_{e2}$ is always negative. Since the power angle is generally in $(-\pi/2, \pi/2)$, $D_{e1}$ is nonnegative. However, under the weak-grid condition and high-power operation mode, the operation power angle is close to $\pi/2$, and $\cos\delta_{pll}$ is also close to zero. Therefore, no matter how the PI parameters of PLL are selected, the positive damping component quickly deteriorates under ultra-weak grid conditions. Based on (1), the transfer function block diagram of a traditional PLL-synchronized inverter is shown in Fig. 2.

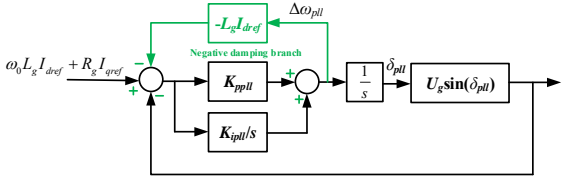

Fig. 2. Transfer function block diagram of a traditional PLL-synchronized inverter.

From Fig. 2, it can be seen that the traditional PLL-synchronized inverter includes a negative damping branch. When the positive damping provided by PLL is sharply reduced under the ultra-weak grid, the negative damping causes synchronization instability.

### III. IMPROVED PLL STRUCTURE

*A. Pure Integral-type PLL with a Damping Branch*

According to the previous analysis results, since the strong nonlinear coupling between the proportional parameter in PLL and the operation power angle, the equivalent damping coefficient sharply drops under ultra-weak grid conditions. Additionally, in the equivalent inertia coefficient, the proportional parameter in PLL is also coupled with the grid impedance and operation current. Therefore, the proportional loop in PLL is canceled, and a pure integral-type PLL (IPLL) is proposed. Also, a positive damping branch is added to compensate for the inherent negative damping. The transfer function block diagram of an IPLL-synchronized inverter is shown in Fig. 3.

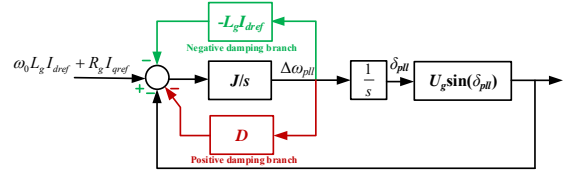

Fig. 3. Transfer function block diagram of the proposed IPLL-synchronized inverter.

According to Fig. 3, the second-order synchronization equation of an IPLL-synchronized inverter is written in (2). It can be found that the equivalent damping coefficient is decoupled with the power angle. The damping coefficient can be always positive with a suitable $D$. Additionally, its equivalent inertia coefficient $1/J$ is a constant. Compared with the traditional PLL-synchronized inverter, the equivalent inertia and damping coefficients of an IPLL-synchronized inverter can be independently designed, respectively.

$$\begin{cases} \dfrac{d\delta_{pll}}{dt} = \omega_{pll} - \omega_g \\ \underbrace{\dfrac{1}{J}}_{J_e} \dfrac{d\omega_{pll}}{dt} = \omega_g L_g I_{dref} + R_g I_{qref} - U_g \sin \delta_{pll} \\ \quad - \underbrace{\left[ D - L_g I_{dref} \right]}_{D_e} (\omega_{pll} - \omega_g) \end{cases} \quad (2)$$

*B. Design of D Coefficient*

From (2), if the feedback gain $D$ satisfies (3) under any conditions, the equivalent damping coefficient can be always positive.

$$D > \max \left[ L_g I_{dref} \right] = \max \left[ \frac{U_q + U_g \sin \delta_{pll} - R_g I_{qref}}{\omega_0} \right] \quad (3)$$

In (3), $R_g$ is small, and the reactive current is not supported by the grid-following inverter. Moreover, considering the ultra-weak grid conditions and high-power operation mode, the operation power angle is close to 90°. Therefore, (3) is rewritten as follows.

$$D > \max \left[ \frac{U_g}{\omega_0} \right] \quad (4)$$

Considering the fluctuation of grid voltage, a conservative design is $D = kU_g/\omega_0$, and $k$ is the voltage fluctuation coefficient, which can keep the damping coefficient always positive under different grid conditions. If the equilibrium point exists, the IPLL-synchronized inverter is always synchronously stable.

According to Fig. 3, the small-signal transfer function at a fixed equilibrium point $\delta_0$ is expressed as follows.

$$H(s) = \frac{1}{J_e s^2 + D_e s + U_g \cos \delta_0} \quad (5)$$

The eigenvalues of (5) are expressed as

$$\lambda_{1,2} = \frac{-D_e \pm \sqrt{D_e^2 - 4 J_e U_g \cos \delta_0}}{2 J_e} \quad (6)$$

Therefore, to avoid overdamping, (6) should be a pair of conjugate roots. Therefore, the maximum value of $D_e$ is limited by (7).

$$D_e < 2\sqrt{J_e V_g \cos \delta_0} \qquad (7)$$

Also, combined with the grid impedance measurement, the adaptive damping coefficient can be easily achieved [9].

*C. Comparison of Damping Coefficients between PLL-synchronized Inverter and IPLL-synchronized Inverter*

In order to fairly compare the PLL-synchronized inverter and IPLL-synchronized inverter, the damping and inertia coefficients are set as the same values under $L_g$=4.1mH (SCR=3). The relationships between parameters in the PLL and IPLL are expressed in (8). With increasing the grid impedance, the damping coefficients of the PLL-synchronized inverter and IPLL-synchronized inverter are shown in Fig. 4.

$$K_{ppll} = \frac{DK_{ipll}}{U_g \cos \delta_0}, K_{ipll} = \frac{JU_g \cos \delta_0}{U_g \cos \delta_0 + JDL_g I_{dref}} \qquad (8)$$

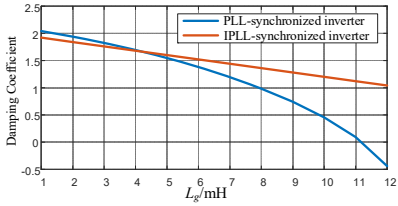

Fig. 4. Damping coefficients of the PLL-synchronized inverter and the proposed IPLL-synchronized inverter.

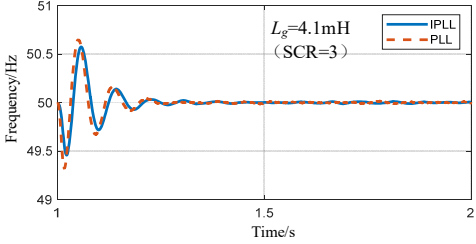

Fig. 5. Simulation results under $L_g$=4.1mH (SCR=3).

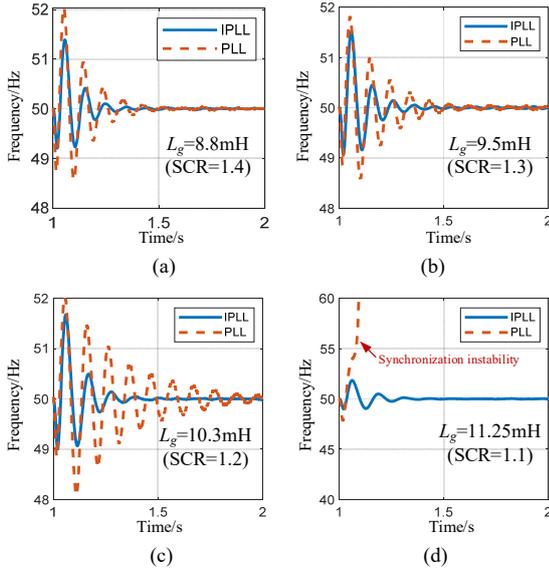

Fig. 6. Time-domain response waveforms under different grid strengths. a) $L_g$=8.8mH. b) $L_g$=9.5mH. c) $L_g$=10.3mH. d) $L_g$=11.25mH.

From Fig.4, it can be seen that the damping coefficients of the PLL-synchronized inverter and IPLL-synchronized inverter are both good under the grid impedance range of 1-7mH. However, with increasing $L_g$, the operation power angle is close to π/2. Since the nonlinearity of the cosine function, the damping coefficient of the PLL-synchronized inverter sharply deteriorates under the grid impedance range of 10-12mH. However, in the IPLL-synchronized inverter, the damping coefficient is not affected by the power angle and is linearly reduced. The damping coefficient is still at a good level even in ultra-weak grid conditions.

## IV. SIMULATION VERIFICATION

An electromagnetic transient time-domain simulation model in Matlab/Simulink is used to verify the correctness of the proposed IPLL-synchronized inverter. The grid voltage is set as $U_g$=311V, and the rated active current of the inverter is set as $I_{dref}$=80A. According to previous analysis, the parameters of IPLL are set as $J$=20 and $D$=2. Under $L_g$=4.1mH (SCR=3), the PI parameters in traditional PLL are $K_{ppll}$=0.1305 and $K_{ipll}$=19.144. Under this condition, the IPLL-synchronized inverter and PLL-synchronized inverter have the same damping coefficient and inertia coefficient. The simulation results are shown in Fig. 5.

From Fig. 5, the IPLL-synchronized inverter and PLL-synchronized inverter have similar time-domain responses. With increasing $L_g$, the time-domain response waveforms are shown in Fig. 6. From Fig. 6, it can be seen that the IPLL-synchronized inverter can stably operate in all grid strengths, and the damping performance is still good under all conditions. As for the PLL-synchronized inverter, the damping performance deteriorates under ultra-weak grid conditions. Under $L_g$=11.25mH, the damping coefficient is negative, and the inverter loses the synchronization stability.

## V. CONCLUSION

In this letter, a pure integral-type PLL is proposed. The equivalent damping coefficient of an IPLL-synchronized inverter is decoupled with the steady-state power angle. Compared with the traditional PLL-synchronized inverter, the IPLL-synchronized inverter can keep a good damping performance even when the power angle is close to 90°. As a result, the IPLL-synchronized inverter can stably operate under an ultra-weak grid when the equilibrium point exists. Additionally, the proposed IPLL is still a second-order system, and its parameters are easily tuned.

Following Inverter Under Weak Grid," *IEEE Trans. Power Electron.,* vol. 37, no. 4, pp. 4091-4104, 2022.